\documentclass[prb,aps,twocolumn,showpacs]{revtex4}
\usepackage{amsmath}
\usepackage{epsfig}

\newcommand{\pr} {Phys. Rev.}

\newcommand{\ijqc} {Int. J. Quant. Chem.}

\newcommand{\zpb} {Z. Physik B}

\newcommand{\xc} {exchange-correlation}

\newcommand{\HF} {Hartree-Fock}

\newcommand{\intsh}{\int \!}
\newcommand{\vecr} {{\mathbf r}}
\newcommand{\veck} {{\mathbf k}}
\newcommand{\vecq} {{\mathbf q}}
\newcommand{\vecG} {{\mathbf G}}
\newcommand{\vecp} {{\mathbf p}}

\newcommand{\Exc} {E_{\rm xc}}

\newcommand{\vTC} {v_{\rm TC}}
\newcommand{\vKS} {v_{\rm KS}}
\newcommand{\vext} {v_{\rm ext}}
\newcommand{\vH} {v_{\rm H}}
\newcommand{\vxc} {v_{\rm xc}}

\newcommand{\dvTC} {\delta \vTC}
\newcommand{\dvKS} {\delta \vKS}
\newcommand{\dvext} {\delta \vext}
\newcommand{\dvH} {\delta \vH}
\newcommand{\dvxc} {\delta \vxc}
\newcommand{\dn} {\delta n}

\newcommand{\FH} {F_{\rm H}}

\newcommand{\Fxc} {F_{\rm xc}}
\newcommand{\Fx} {F_{\rm x}}
\newcommand{\Hx} {H_{\rm x}}
\newcommand{\uC} {w_C}

\newcommand{\chitilde} {\widetilde{\chi}}

\newcommand{\ouC} {\widehat{w}_C}
\newcommand{\ovxnl} {\widehat{\Sigma}_x}
\newcommand{\ovx} {\widehat{v}_x}

\newcommand{\ak} {a\veck}
\newcommand{\akp} {a\veck'}
\newcommand{\akq} {a\veck + \vecq}
\newcommand{\bk} {b\veck}
\newcommand{\bkp} {b\veck'}
\newcommand{\bkq} {b\veck + \vecq}
\newcommand{\bkpq} {b\veck' + \vecq}
\newcommand{\sk} {s\veck}
\newcommand{\skq} {s\veck+\vecq}
\newcommand{\tk} {t\veck}
\newcommand{\tkp} {t\veck'}
\newcommand{\tkq} {t\veck + \vecq}
\newcommand{\tkpq} {t\veck'+\vecq}

\newcommand{\eak} {\epsilon_{\ak}}

\newcommand{\eskq} {\epsilon_{\skq}}
\newcommand{\ebk} {\epsilon_{\bk}}
\newcommand{\ebkp} {\epsilon_{\bkp}}
\newcommand{\etkq} {\epsilon_{\tkq}}
\newcommand{\etkpq} {\epsilon_{\tkpq}}

\newcommand{\ask} {as\veck}

\newcommand{\btk} {bt\veck}

\newcommand{\wwpid} {\omega + i \delta}
\newcommand{\wwmid} {\omega - i \delta}

\newcommand{\lak} {\langle \ak |}
\newcommand{\lakq} {\langle \akq |}
\newcommand{\rak} {| \ak \rangle}
\newcommand{\rakq} {| \akq \rangle}
\newcommand{\lsk} {\langle \sk |}
\newcommand{\lskq} {\langle \skq |}
\newcommand{\rsk} {| \sk \rangle}
\newcommand{\rskq} {| \skq \rangle}
\newcommand{\lbk} {\langle \bk |}
\newcommand{\lbkq} {\langle \bkq|}
\newcommand{\lbkpq} {\langle \bkpq|}
\newcommand{\rbk} {| \bk \rangle}
\newcommand{\rbkp} {| \bkp\rangle}
\newcommand{\rbkq} {| \bkq \rangle}
\newcommand{\ltk} {\langle \tk |}
\newcommand{\ltkq} {\langle \tkq|}
\newcommand{\ltkpq} {\langle \tkpq|}
\newcommand{\rtk} {| \tk \rangle}
\newcommand{\rtkp} {| \tkp\rangle}
\newcommand{\rtkq} {| \tkq\rangle}

\newcommand{\qzero} {q \rightarrow 0}

\begin{document}

\title{Exact Kohn-Sham exchange kernel for insulators and
its long-wavelength behavior}

\author{Yong-Hoon Kim}
\altaffiliation{Present address:
    Materials and Process Simulation Center (139-74),
    California Institute of Technology,
    Pasadena, CA 91125-7400}
\author{Andreas G{\"o}rling}
\affiliation{Lehrstuhl f{\"u}r Theoretische Chemie,
    Technische Universit{\"a}t M{\"u}nchen,
    D-85748 Garching, Germany}

\date{\today}

\begin{abstract}
We present an exact expression for the frequency-dependent
Kohn-Sham exact-exchange (EXX) kernel for periodic insulators,
which can be employed for the calculation of electronic response
properties within time-dependent (TD) density-functional theory.
It is shown that the EXX kernel has a long-wavelength divergence
behavior of the exact full exchange-correlation kernel and thus
rectifies one serious shortcoming of the adiabatic local-density
approximation and generalized-gradient approximations kernels. A
comparison between the TDEXX and the
GW-approximation-Bethe-Salpeter-equation approach is also made.
\end{abstract}


\maketitle

\section{Introduction}
\label{sec:intro}

Time-dependent density-functional theory
(TDDFT)~\cite{Gross96,Casida95} is an attractive first-principles
formalism for the calculation of electronic response properties.
Thanks to its simplicity, applications to quantum
wells~\cite{Ando77} and atoms~\cite{Zangwill80} have appeared long
before its formal justification,~\cite{Runge84} and currently its
use is ubiquitous in a wide range of \textit{ab initio}
computations.  A crucial ingredient of successful TDDFT
applications is the approximation to the {\em dynamic} \xc\
kernel,
\begin{equation}
\label{eq:Fx}
 \Fxc(\vecr,\vecr';t-t')
    \equiv \frac{\delta \vxc(\vecr;t)}{\dn(\vecr';t')},
\end{equation}
which together with the Hartree kernel~\cite{AU}
$\FH(\vecr,\vecr') = \uC(\vecr,\vecr') \equiv 1 / |\vecr -
\vecr'|$ completely determines the two-particle interaction
effects.  For $\Fxc$, the adiabatic local-density approximation
(LDA) kernel,
\begin{equation}
\label{eq:Fx-LDA}
 \Fxc^{LDA}(\vecr,\vecr';t-t')
    = \delta(t-t') \delta(\vecr-\vecr') \frac{d \vxc^{LDA}[n(\vecr)]}{d
    n(\vecr)},
\end{equation}
has been almost exclusively adopted in practical calculations.
However, the scope of the LDA kernel has been rather limited for
infinite periodic solids due to its deficiencies, and in
particular its incorrect non-divergent long-wavelength behavior
for insulators has been emphasized as a primary defect in recent
years.~\cite{Aulbur96,Ghosez97,Reining01}  This shortcoming of the
LDA kernel shows up, e.g., in its incapability of describing
excitonic effects in absorption spectra of solids.  The semilocal
generalized gradient approximations (GGA) kernel does not improve
over the LDA one in this case, and the task of developing a more
accurate approximate \xc\ kernel remains as a challenging task for
the TDDFT study of solids.

Indeed deficiencies of the LDA and GGA appear already at the level
of the {\em static} \xc\ energy functional $E_{xc}[n]$ and the
\xc\ potential $\vxc(\vecr) \equiv \delta \Exc[n] / \dn(\vecr)$.
For instance, the LDA and GGA $E_{xc}[n]$ inherently fails to
describe the quasi-two-dimensional electron gas due to their
(semi)local nature.~\cite{2D-egas} For $\vxc$ and the
corresponding Kohn-Sham (KS) eigenvalues, the LDA and GGA $\vxc$
incorrectly decays exponentially rather than as $-1/r$ for
localized systems, and consequently their highest occupied orbital
energies are too high and unoccupied orbital energies do not
exhibit Rydberg series. For solids, the LDA band gaps are too
small and this behavior is again not corrected by the GGA. In
fact, even the exact KS gap does not equal the experimental band
gap but differs by the discontinuity of the \xc\
potential.~\cite{discontinuity} However, recent theoretical and
numerical studies suggested that the KS equation for $N$ electrons
corresponds to the Dyson equation where the reference ground state
is chosen with $N-1$ electrons,~\cite{Umrigar98} and accordingly
unoccupied orbitals in the KS calculations should give a good
description of excitations of the $N$-electron system with fixed
particle number. This is in accordance with the perturbation
theory along the adiabatic connection~\cite{Gorling96b} which
finds that differences of KS eigenvalues represent the leading
term in the expansion of excitation energies.

In this regard, recent development of the KS exact-exchange (EXX)
method, which treats the \xc\ energy functional exactly in leading
order in the electron-electron interaction,  provides an
interesting opportunity. Self-interaction-free, nonlocal EXX
schemes give not only realistic exchange potentials and KS
eigenvalue spectra for
molecules~\cite{Grabo97,KLI99,Gorling99,Ivanov99,DellaSala01} but
also band structures of semiconductors in good agreement with
experiments.~\cite{Bylander95,Stadele97}  We have recently shown
that the EXX orbitals and eigenvalues at the one-particle level
without any previously applied post-DFT modification such as the
quasiparticle shift~\cite{Levine89} indeed give a very good
description of the absorption spectrum of semiconductors with the
exception of excitonic features resulting from two-particle
interactions, and argued that it is another evidence of the
above-described picture of ``KS quasiparticles''.~\cite{DFT01}

In view of the encouraging performance of the EXX method, we
present in this work an exact expression of the EXX kernel
$\Fx^{EXX}$ for periodic insulators which can be employed for
calculations of electronic linear response properties within
TDDFT.  It will be shown that the EXX kernel, unlike the LDA and
GGA kernels, exhibits a long-wavelength behavior of the exact
$\Fxc$ which is particularly important for the study of electronic
excitations in infinite solids.  This behavior of the EXX kernel
has been previously claimed by Goshez \textit{et al.}  based on a
plausibility argument,~\cite{Ghosez97} and here we explicitly
prove this using our exact formula.

\section{Long-wavelength behavior of the exchange-correlation kernel
for insulators}
\label{sec:q-zero}

We first establish the definitions and notations of various
quantities of interest and derive the long-wavelength behavior of
the exact $\Fxc$.~\cite{Ghosez97} The full linear density response
matrix $\chi$ describes the response of the first-order (number)
density change $\dn$ for the given bare dynamic perturbation
$\dvext$,
\begin{equation}
\label{eq:dn-chi}
 \dn(\vecG,\vecq;\omega) = \sum_{\vecG'} \chi(\vecG,\vecG',\vecq,\omega)
    \, \dvext(\vecG',\vecq;\omega).
\end{equation}
We chose to work in the reciprocal space and the frequency domain,
e.g., $\chi(\vecG,\vecG',\vecq;\omega)$ is a matrix in the
reciprocal-space lattice vectors $\vecG$ and $\vecG'$ for the
given wave-vector $\vecq$ and frequency $\omega$. Analysis of the
$\qzero$ behavior of the ``head'' ($\vecG = \vecG' = 0$), ``wing''
($\vecG = 0$ and $\vecG' \neq 0$ or vice versa), and ``body''
($\vecG \neq 0$ and $\vecG' \neq 0$) elements of $\chi$ and other
related matrices appearing below is an important discussion point
throughout the paper. From now on, we will adopt the matrix
notation and $\vecG$ dependence will be assumed unless explicitly
stated otherwise. Within TDDFT, $\dn$ is expressed in terms of the
dynamic  linear response matrix $\chi_0$ and the first-order
change of the effective KS potential $\dvKS$,
\begin{equation}
\label{eq:dn-chi0}
 \dn(\vecq;\omega) = \chi_0(\vecq;\omega) \, \dvKS(\vecq;\omega),
\end{equation}
where $\dvKS$ is composed of the external perturbation $\dvext$
and the resulting change in the Hartree potential $\dvH$ and the
\xc\ potential $\dvxc$,
\begin{equation}
\label{eq:dvKS}
\begin{split}
  \dvKS(\vecq;\omega)
    & = \dvext(\vecq;\omega) + \dvH(\vecq;\omega) + \dvxc(\vecq;\omega)\\
    & = \dvext(\vecq;\omega) + \bigl[ \FH(\vecq) + \Fxc(\vecq;\omega) \bigr] \,
    \dn(\vecq;\omega),
\end{split}
\end{equation}
with $\FH(\vecG,\vecG',\vecq) = \delta_{\vecG,\vecG'}
4 \pi / |\vecq + \vecG|^2
$. Then, from Eqs. (\ref{eq:dn-chi}),
(\ref{eq:dn-chi0}), and (\ref{eq:dvKS}), one obtains
\begin{equation}
 \chi_0^{-1}(\vecq;\omega) =
    \chi^{-1}(\vecq;\omega) + \FH(\vecq) + \Fxc(\vecq;\omega),
\end{equation}
which shows that $\chi$ is completely determined once $\chi_0$ and
$\Fxc$ are given.

For further consideration of the $\qzero$ behavior of $\Fxc$, it
is convenient to introduce the ``proper'' part of $\chi$,
$\chitilde$, defined through~\cite{Pick70,Ghosez97}
\begin{equation}
\label{eq:dn-chitilde}
 \dn(\vecq;\omega) = \chitilde(\vecq;\omega) \, \dvTC(\vecq;\omega),
\end{equation}
where $\dvTC$ is the change of the test-charge potential,
\begin{equation}
\label{eq:dvtc}
  \dvTC(\vecq;\omega) \equiv \dvext(\vecq;\omega) + \FH(\vecq) \,
\dn(\vecq;\omega),
\end{equation}
and thus relates with the full response matrix $\chi$ as
\begin{equation}
\label{eq:chitilde-chi}
 \chitilde^{-1}(\vecq;\omega)
    = \chi^{-1}(\vecq;\omega) + \FH(\vecq),
\end{equation}
or with the KS response matrix $\chi_0$ as
\begin{equation}
\label{eq:chi0-chitilde}
 \chi_0^{-1}(\vecq;\omega)
    = \chitilde^{-1}(\vecq;\omega) + \Fxc(\vecq;\omega).
\end{equation}
The linear response matrix of the {\em non-interacting KS} system
$\chi_0$ and the proper part of that of the {\em real interacting}
system $\chitilde$ are known to have the following similar
$\qzero$ behavior~\cite{Adler62,Pick70} (assuming that both the KS
system and the real system are insulating~\cite{Ghosez97}):
\begin{equation}
\label{eq:chi-q0}
  \chi_0 =
    \begin{bmatrix}
    q^2 \, \chi_0^{00} & q \, \chi_0^{01} \\
    q   \, \chi_0^{10} &      \chi_0^{11}
    \end{bmatrix}; \;
  \chitilde =
    \begin{bmatrix}
    q^2 \, \chitilde^{00} & q \, \chitilde^{01} \\
    q   \, \chitilde^{10} &      \chitilde^{11}
    \end{bmatrix},
\end{equation}
where we used the notation that $\xi^{00}$, $\xi^{01/10}$, and
$\xi^{11}$ denote the head of a matrix $\xi$ divided by $q^2$, its
wings divided by $q$, and its body, respectively. The quantities
$\chi_0^{00}$, $\chi_0^{10}$, $\chi_0^{01}$, and $\chi_0^{11}$ as
well as $\chitilde^{00}$, $\chitilde^{10}$, $\chitilde^{01}$, and
$\chitilde^{11}$ all consist of a leading {\em
$\vecq$-independent} term and contributions of higher order in $q$
which vanish in the limit $\qzero$. Then, from Eqs.
(\ref{eq:chi0-chitilde}) and (\ref{eq:chi-q0}), one can deduce
that the head and wings of $\Fxc$ have the following divergent
$\qzero$ behavior~\cite{Ghosez97} (assuming that there exists no
fortuitous cancellation between $\chi_0^{-1}$ and
$\chitilde^{-1}$),
\begin{equation}
 \label{eq:fxc-q0}
 \Fxc =
    \begin{bmatrix}
    \Fxc^{00}/q^2 &  \Fxc^{01}/q \\
    \Fxc^{10}/q &        \Fxc^{11}
    \end{bmatrix},
\end{equation}
In Eq.(\ref{eq:fxc-q0}), $\Fxc^{00}$, $\Fxc^{10}$, $\Fxc^{01}$,
and $\Fxc^{11}$ again contain a leading {\em $\vecq$-independent}
term and contributions of higher order in $q$ which vanish for
$\qzero$. The head and wings of adiabatic LDA and GGA kernels, on
the other hand, are independent of $q$ and thus are incorrectly
non-divergent for $\qzero$:
\begin{equation}
 \Fxc^{LDA/GGA} =
    \begin{bmatrix}
    \Fxc^{LDA/GGA,00} & \Fxc^{LDA/GGA,01} \\
    \Fxc^{LDA/GGA,10} & \Fxc^{LDA/GGA,11}
    \end{bmatrix}.
\end{equation}
This defect is a serious problem not only from a theoretical
viewpoint but also for practical purposes because the head and
wings of the \xc\ kernel can affect the the macroscopic dielectric
function in leading order.~\cite{Ghosez97} For example, it has
been recently shown that they play a crucial role for the proper
treatment of excitonic effects in the calculation of optical
spectra.~\cite{Reining01}

\section{Exact-exchange kernel and its long-wavelength behavior for
insulators}
\label{sec:Fx-EXX}

Deficiencies of the LDA and GGA kernels discussed above represent
a major problem from the theoretical and calculational point of
view which could not be overcome so far. To ameliorate the
situation we propose to adopt the EXX kernel.  An exact expression
of the EXX kernel has been previously derived by one of us for
localized systems for the case of real-valued
orbitals.~\cite{Gorling98} For periodic solids, we need to
generalize this expression to complex orbitals and have to
consider the dependence on wave vectors $\veck$ and $\vecq$. This
leads to
\begin{equation}
\begin{split}
 \label{eq:Fx-EXX}
\Fx^{EXX} & (\vecG,\vecG',\vecq;\omega) = \sum_{\vecG_1,\vecG_2}
    \chi_0^{-1}(\vecG,\vecG_1,\vecq; \omega) \\
    & \times \Hx(\vecG_1,\vecG_2,\vecq;\omega)
    \, \chi_0^{-1}(\vecG_2,\vecG',\vecq;\omega),
\end{split}
\end{equation}
where the EXX kernel ``core'' $\Hx$ is composed of the following
contributions (We assume that $\dvext(\vecq;\omega)$ and other
quantities have the time-dependence $e^{-i\omega t} e^{\delta t}$,
where $\delta \rightarrow 0^+$ is a convergence factor.):
\begin{widetext}
\begin{equation}
\label{eq:Hx-1}
\begin{split}
\Hx^1(\vecG,\vecG',\vecq;\omega) \equiv
  \!- \frac{2}{\Omega} \!\sum_{\ask}\!\sum_{\btk'} & \Bigl[
     \frac{\lak e^{-i(\vecq+\vecG) \cdot \vecr} \rskq
     \langle \skq; \bkp |\ouC| \tkpq; \ak \rangle
     \ltkpq e^{i(\vecq+\vecG') \cdot \vecr} \rbkp}
         {(\eak - \eskq + \wwpid)(\ebkp - \etkpq + \wwpid)} \\
 + \ &\frac{\lsk e^{-i(\vecq+\vecG) \cdot \vecr} \rakq
     \langle \akq ; \tkp |\ouC| \bkpq ; \sk \rangle
     \lbkpq e^{i(\vecq+\vecG') \cdot \vecr} \rtkp}
         {(\eak - \eskq - \wwmid)(\ebkp - \etkpq - \wwmid)}
 \Bigr],
\end{split}
\end{equation}
\begin{equation}
\label{eq:Hx-2}
\begin{split}
\Hx^2(\vecG,\vecG',\vecq;\omega) \equiv
 \!- \frac{2}{\Omega} \!\sum_{\ask}\!\sum_{\btk'} & \Bigl[
    \frac{\lak e^{-i(\vecq+\vecG) \cdot \vecr} \rskq
     \langle \skq; \tkp |\ouC| \bkpq; \ak \rangle
     \lbkpq e^{i(\vecq+\vecG') \cdot \vecr} \rtkp}
         {(\eak - \eskq + \wwpid)(\ebkp - \etkpq - \wwmid)} \\
 + \ &\frac{\lsk e^{-i(\vecq+\vecG) \cdot \vecr} \rakq
     \langle \akq; \bkp |\ouC| \tkpq; \sk \rangle
     \ltkpq e^{i(\vecq+\vecG') \cdot \vecr} \rbkp}
         {(\eak - \eskq - \wwmid)(\ebkp - \etkpq + \wwpid)}
 \Bigr],
\end{split}
\end{equation}
\begin{equation}
\label{eq:Hx-3}
\begin{split}
\Hx^3(\vecG,\vecG',\vecq;\omega) \equiv
 \!- \frac{2}{\Omega} \!\sum_{abs\veck} \Bigl[
    &\frac{ \lak e^{-i(\vecq+\vecG) \cdot \vecr} \rskq
     \lbk \ovxnl - \ovx \rak
     \lskq e^{i(\vecq+\vecG') \cdot \vecr} \rbk}
    {(\eak - \eskq + \wwpid)(\ebk - \eskq + \wwpid)}\\
 + \    &\frac{ \lsk e^{-i(\vecq+\vecG) \cdot \vecr} \rakq
     \lakq \ovxnl - \ovx \rbkq
     \lbkq e^{i(\vecq+\vecG') \cdot \vecr} \rsk}
    {(\eak - \eskq - \wwmid)(\ebk - \eskq - \wwmid)}
 \Bigr]\\
 + \frac{2}{\Omega} \!\sum_{ast\veck} \Bigl[
    &\frac{ \lak e^{-i(\vecq+\vecG) \cdot \vecr} \rskq
        \lskq \ovxnl - \ovx \rtkq
        \ltkq e^{i(\vecq+\vecG') \cdot \vecr} \rak}
    {(\eak - \eskq + \wwpid)(\eak - \etkq + \wwpid)} \\
 + \    &\frac{\lsk e^{-i(\vecq+\vecG) \cdot \vecr} \rakq
        \ltk \ovxnl - \ovx \rsk
        \lakq e^{i(\vecq+\vecG') \cdot \vecr} \rtk}
    {(\eak - \eskq - \wwmid)(\eak - \etkq - \wwmid)}
 \Bigr],
\end{split}
\end{equation}
and
\begin{equation}
\label{eq:Hx-4}
\begin{split}
\Hx^4(\vecG,\vecG',\vecq;\omega) \equiv
 \!- \frac{2}{\Omega} \!\sum_{abs\veck} \Bigl[
     &\frac{ \lbk e^{-i(\vecq+\vecG) \cdot \vecr} \rskq
     \lskq \ovxnl - \ovx \rakq
     \lakq e^{i(\vecq+\vecG') \cdot \vecr} \rbk}
    {(\ebk - \eskq + \wwpid) (\eak - \eskq)} \\
 + \    &\frac{ \lbk e^{-i(\vecq+\vecG) \cdot \vecr} \rakq
     \lakq \ovxnl - \ovx \rskq
     \lskq e^{i(\vecq+\vecG') \cdot \vecr} \rbk}
    {(\eak - \eskq) (\ebk - \eskq + \wwpid)} \\
 + \    &\frac{ \lsk e^{-i(\vecq+\vecG) \cdot \vecr} \rbkq
     \lak \ovxnl - \ovx \rsk
     \lbkq e^{i(\vecq+\vecG') \cdot \vecr} \rak}
    {(\ebk - \eskq - \wwmid) (\eak - \eskq)} \\
 + \    &\frac{ \lak e^{-i(\vecq+\vecG) \cdot \vecr} \rbkq
     \lsk \ovxnl - \ovx \rak
     \lbkq e^{i(\vecq+\vecG') \cdot \vecr} \rsk}
    {(\eak - \eskq) (\ebk - \eskq - \wwmid)}
 \Bigr]\\
 + \frac{2}{\Omega} \!\sum_{ast\veck} \Bigl[
        &\frac{\lak e^{-i(\vecq+\vecG) \cdot \vecr} \rtkq
        \lsk \ovxnl - \ovx \rak
        \ltkq e^{i(\vecq+\vecG') \cdot \vecr} \rsk}
    {(\eak - \etkq + \wwpid) (\eak - \eskq)} \\
 + \    &\frac{\lsk e^{-i(\vecq+\vecG) \cdot \vecr} \rtkq
        \lak \ovxnl - \ovx \rsk
        \ltkq e^{i(\vecq+\vecG') \cdot \vecr} \rak}
    {(\eak - \eskq) (\eak - \etkq + \wwpid)} \\
 + \    &\frac{\ltk e^{-i(\vecq+\vecG) \cdot \vecr} \rakq
        \lakq \ovxnl - \ovx \rskq
        \lskq e^{i(\vecq+\vecG') \cdot \vecr} \rtk}
    {(\eak - \etkq - \wwmid) (\eak - \eskq)} \\
 + \    &\frac{\ltk e^{-i(\vecq+\vecG) \cdot \vecr} \rskq
        \lskq \ovxnl - \ovx \rakq
        \lakq e^{i(\vecq+\vecG') \cdot \vecr} \rtk}
    {(\eak - \eskq) (\eak - \etkq - \wwmid)}
 \Bigr].
\end{split}
\end{equation}
In Eqs.(\ref{eq:Hx-1})-(\ref{eq:Hx-4}), 2 is the spin factor,
$\Omega$ is the crystal volume, $\{a,b\}$ are valence bands,
$\{s,t\}$ are conduction bands, $\langle i\veck+\vecq;j\veck'|
\ouC |l\veck'+\vecq;m\veck \rangle$ are four-index integrals
defined as
\begin{equation}
\label{eq:4pt-integrals}
\begin{split}
 \langle i\veck+\vecq ; j\veck' |\ouC| l\veck'+\vecq; m\veck \rangle
    & \equiv \intsh d\vecr \! \intsh d\vecr' \
    \frac{\phi_{i\veck+\vecq}^*(\vecr) \phi_{j\veck'}^*(\vecr')
        \phi_{l\veck'+\vecq}(\vecr) \phi_{m\veck}(\vecr')}
    {|\vecr - \vecr'|} \\
    & = \frac{4 \pi}{\Omega} \sum_{\vecG}
    \frac{ \langle i\veck+\vecq| e^{i (\vecG + \veck - \veck') \cdot \vecr}
            | l\veck'+\vecq \rangle
        \langle j\veck' | e^{-i (\vecG + \veck -\veck') \cdot \vecr'}
            | m \veck \rangle}
        {|\vecG + \veck - \veck'|^2},
\end{split}
\end{equation}
$\ovxnl$ is a nonlocal orbital-dependent exchange operator of the
form of the \HF\ exchange operator but constructed with the KS
orbitals $\phi_a$,
\begin{equation}
\label{eq:nonlocal-v}
\begin{split}
 \langle i\veck+\vecq | \ovxnl | j\veck+\vecq \rangle
 & \equiv - \intsh d\vecr \! \intsh d\vecr' \phi^*_{i\veck+\vecq}(\vecr)
    \sum_{a\veck'} \frac{\phi_{\akp}(\vecr) \phi_{\akp}^*(\vecr')}
    {|\vecr-\vecr'|}\phi_{j\veck+\vecq}(\vecr')\\
 & = - \sum_{a\veck'} \langle i\veck+\vecq; a\veck' |\ouC| a\veck'; j\veck+\vecq \rangle\\
 & = - \frac{4 \pi}{\Omega} \sum_{a\veck'\vecG}
    \frac{ \langle i\veck+\vecq| e^{i (\vecG + \veck - \veck'+\vecq) \cdot \vecr}
            | a\veck' \rangle
        \langle a\veck' | e^{-i (\vecG + \veck -\veck'+\vecq) \cdot \vecr'}
            | j \veck+\vecq \rangle}
        {|\vecG + \veck - \veck' + \vecq|^2},
\end{split}
\end{equation}
\end{widetext}
and $\ovx$ is generated by the local multiplicative EXX KS
potential $v_x(\vecr)$.

Compared with the LDA (or GGA) kernel which is (semi)local in real
space and frequency-independent [Eq.~(\ref{eq:Fx-LDA})], which
results in a reciprocal-representation independent of $\vecq$ and
$\omega$, $\Fxc^{LDA/GGA}(\vecG,\vecG',\vecq;\omega) =
\Fxc^{LDA/GGA}(\vecG-\vecG')$, $\Fx^{EXX}$ is fully nonlocal in
real space and depends explicitly on the frequency. We now show
that $\Fx^{EXX}$ has a $\qzero$ behavior as the exact $\Fxc$. By
expanding orbitals $\phi_{j\veck+\vecq}$ in terms of the orbitals
$\phi_{i\veck}$ employing perturbation
theory,~\cite{Adler62,Pick70}
\begin{equation}
 \phi_{j \veck+\vecq} =
    \phi_{j \veck} +
    \sum_{i \neq i} \phi_{i\veck}
    \frac{ \vecq \cdot \langle i\veck| \vecp | j\veck \rangle}
        {\epsilon_{i\veck} - \epsilon_{j\veck}},
\end{equation}
we express various matrix elements of
Eqs.(\ref{eq:Hx-1})$-$(\ref{eq:Hx-4}) in power series in $\vecq$
and keep only the leading non-vanishing terms. Then, one can first
observe that $\langle i\veck;j\veck' | l\veck';m\veck \rangle$ and
$\langle i\veck | \ovxnl - \ovx | j\veck \rangle$ are the leading
order terms in $q$ of $\langle i\veck+\vecq;j\veck' |
l\veck'+\vecq;m\veck \rangle$ and $\langle i\veck+\vecq | \ovxnl -
\ovx | j\veck+\vecq \rangle$ and that consequently the
$\vecq$-dependence can be ignored for $\qzero$ in the inner matrix
elements of Eqs. (\ref{eq:Hx-1})$-$(\ref{eq:Hx-4}). One might
notice that the inner matrix elements $\langle i{\bf 0}+\vecq;
j{\bf 0} | i{\bf 0}+\vecq; j{\bf 0} \rangle$ in $\Hx^1$ contain a
singular contribution, the term with $\vecG=0$ in Eq.
(\ref{eq:4pt-integrals}). However, the same singularities with the
opposite sign arise in the matrix elements $\langle i\veck+\vecq |
\ovxnl | j\veck+\vecq \rangle$ of the first two contributions of
$\Hx^3$, the terms with $\vecG=0$ and $\veck'=\veck+\vecq$ in Eq.
(\ref{eq:nonlocal-v}). So the Coulomb singularities in the inner
matrix elements of $\Hx^1$ and $\Hx^3$ exactly cancel.

Unlike in the case of inner matrix elements, for the other outer
matrix elements $\langle i\veck | e^{-i(\vecq+\vecG) \cdot \vecr}
| j\veck+\vecq \rangle$ or $\langle i\veck+\vecq |
e^{-i(\vecq+\vecG') \cdot \vecr} | j\veck \rangle$ with $i\ne j$,
$\vecq$-dependent contributions appear in leading order in $q$ for
$\vecG=0$ or $\vecG'=0$ \cite{Adler62}, e.g.,
\begin{equation}
\label{eq:q0-limit}
 \lim_{\qzero} \langle i\veck | e^{-i \vecq \cdot \vecr} |
 j\veck+\vecq \rangle = \vecq \cdot \frac{\langle i\veck | \vecp |
 j\veck \rangle} {\epsilon_{i\veck} - \epsilon_{j\veck}}.
\end{equation}
In $\Hx^4$, however, matrix elements
$\langle i\veck | e^{-i(\vecq+\vecG) \cdot \vecr} |
i\veck+\vecq \rangle$ and $\langle i\veck+\vecq |
e^{-i(\vecq+\vecG') \cdot \vecr} | i\veck \rangle$
are present, for which a leading order term independent of $q$
occurs for $\vecG=0$ or $\vecG'=0$. However,
contributions of such type in the first sum of $\Hx^4$ are
cancelled by corresponding contributions in the second sum.

Due to the cancellations of singularities, $\Hx$ itself is
well-defined, and the $\qzero$ behavior of $\Hx$ can be deduced as
\begin{equation}
\label{eq:Hx-q0}
  \Hx = \begin{bmatrix}
    q^2 \, \Hx^{00} & q \, \Hx^{01} \\
    q   \, \Hx^{10} &      \Hx^{11}
    \end{bmatrix}
\end{equation}
with $\Hx^{00}$, $\Hx^{10}$, $\Hx^{01}$, and $\Hx^{11}$ containing
a leading order term independent of $q$.
Consequently, using Eqs. (\ref{eq:chi-q0}), (\ref{eq:Fx-EXX}), and
(\ref{eq:Hx-q0}), we conclude that $\Fx^{EXX}$ has the
$\qzero$ behavior of the exact $\Fxc$,
\begin{equation}
  \Fx^{EXX} =
    \begin{bmatrix}
    \Fx^{EXX,00}/q^2 & \Fx^{EXX,01}/q \\
    \Fx^{EXX,10}/q &        \Fx^{EXX,11}
    \end{bmatrix}
\end{equation}
with $\Fx^{00}$, $\Fx^{10}$, $\Fx^{01}$, and $\Fx^{11}$, again
containing a $q$-independent leading order term.

\section{Discussion and conclusions}
\label{sec:discussions}

Now we analyze the physical meaning of $\Hx$ and relate it with
the GW approximation (GWA)-Bethe-Salpeter equation (BSE)
approach,~\cite{Rohlfing00} which represents at the moment the
most successful first-principles computational scheme of
electronic excitations in solids.  We start by rewriting
Eq.~(\ref{eq:chi0-chitilde}) as $\chitilde = (1-\chi_0\Fxc)^{-1}
\chi_0$. By first expanding $(1-\chi_0\Fxc)^{-1}$ in a power
series into $1+\chi_0\Fxc+\chi_0\Fxc\chi_0\Fxc+ ...$, next taking
only the first two leading terms of this expansion, and finally
neglecting correlation contributions, $(1-\chi_0\Fxc)^{-1} \approx
1+\chi_0\Fx^{EXX}$, we obtain $\chitilde \approx \chi_0 +
\chi_0\Fx^{EXX}\chi_0$. Thus, identifying $\chi_0\Fx^{EXX}\chi_0$
as $\Hx$ [See Eq.~(\ref{eq:Fx-EXX})], we can interpret $\Hx$ as
the first order correction to $\chi_0$ in $\chitilde$,
\begin{equation}
\label{eq:Hx-meaning}
 \chitilde(\vecq;\omega) \approx \chi_0(\vecq;\omega) +
 \Hx(\vecq;\omega).
\end{equation}
Indeed, $\Hx$ has been recently shown in the many-body
diagrammatic language as the first-order self-energy and vertex
corrections to the irreducible polarizability
$\chitilde$.~\cite{Tokatly01}

\begin{figure*}

\begin{minipage}[H]{0.7\linewidth}
 \epsfig{file=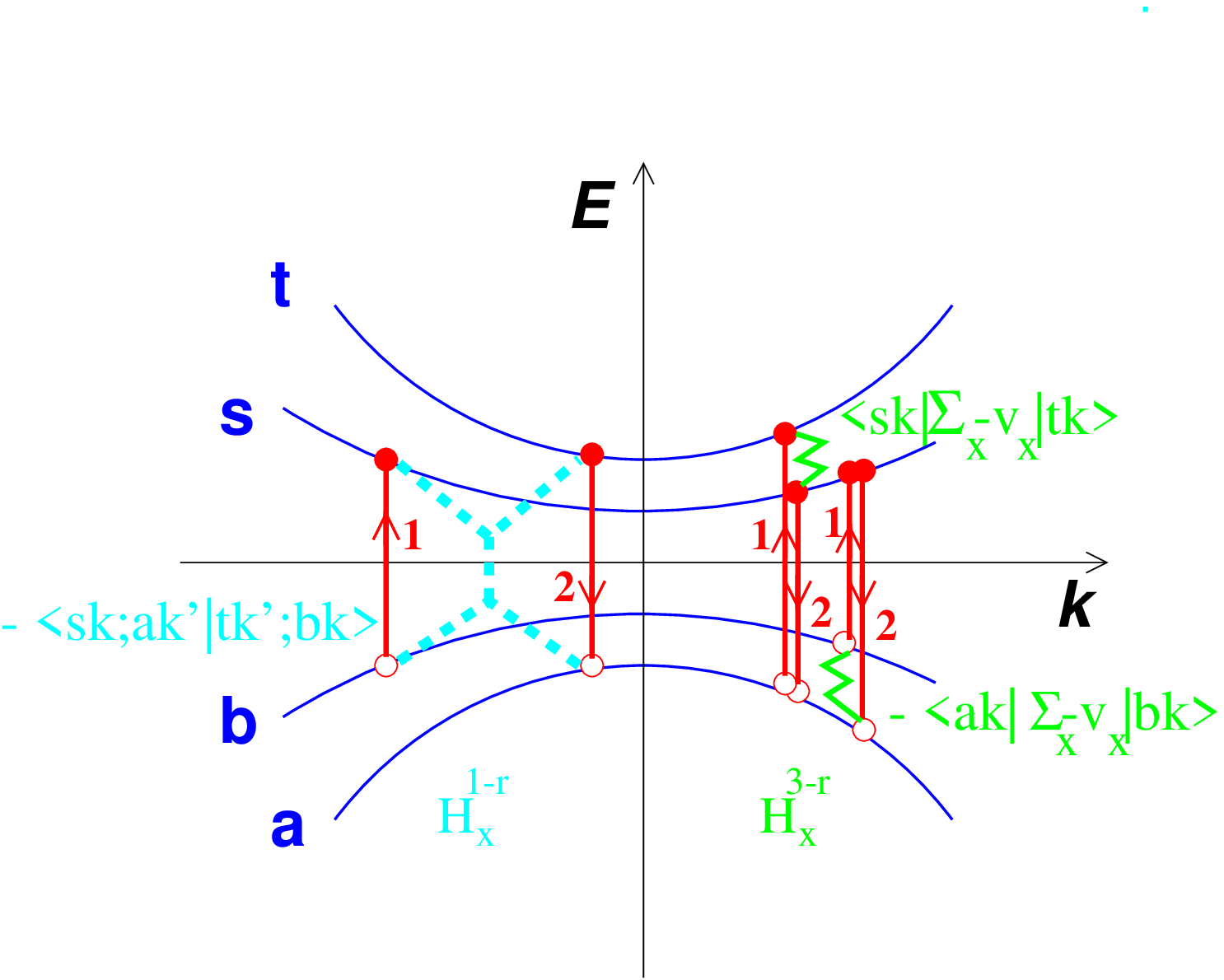,width=\linewidth}
\end{minipage}

\caption{ \label{fig:Hx_resonant} Schematic description of the
resonant contributions to $\Hx$ ($\Hx^{1-r}$ and $\Hx^{3-r}$).
Arrow 1 and 2 represent $\langle s \veck+\vecq | e^{i (\vecq +
\vecG) \cdot \vecr} | a \veck \rangle$ and $\langle a \veck |
e^{-i (\vecq + \vecG) \cdot \vecr} | s \veck+\vecq \rangle$. They
involve the {\em time-sequential} coupling of an electron
excitation from valence $\{a,b\}$ to conduction $\{s,t\}$ bands
(hole $\rightarrow$ electron pair) and a relaxation from
conduction to valence bands (electron $\rightarrow$ hole pair).}
\end{figure*}

The expression of the full $\Hx$ is admittedly quite complicated.
However, we point out that a simplified picture of the important
underlying physical processes within TDEXX can be extracted by
noting that only the first term of $\Hx^1$ ($\Hx^{1-r}$) and the
first and third terms of $\Hx^3$ ($\Hx^{3-r}$) are dominant
contributions at resonant $\omega$. This is schematically depicted
in Fig. \ref{fig:Hx_resonant}.

Note that the above situation is similar to the one that occurs in the
solution of the BSE where the Tamm-Dancoff approximation is
invoked.~\cite{Rohlfing00} In fact, with the EXX kernel, we can
easily make a connection between the TDDFT and the GWA-BSE
approach. Consider calculation of the full response
function~\cite{Reining01} or excitation
energies~\cite{Gonze99,Tokatly01} with TDEXX and GWA-BSE.
Replacing the bare Coulomb interaction with the screened Coulomb
interaction, the resonant terms of $\Hx^3$ effectively shift the
EXX eigenvalue spectrum toward that of the GWA, while the resonant
terms in $\Hx^1$ are the counterparts of those occurring in the
BSE in the Tamm-Dancoff approximation.

In the above comparison, it is interesting to observe that while
the GWA and BSE have a clear hierarchy as the theory of
independent quasiparticle excitations~\cite{quasiptl} and
electron-hole excitations, terms related to both excitation
effects appear within the DFT formulation at the time-dependent
level and the distinction between one- and two-particle
excitations is accordingly rather arbitrary. We should also
mention that the mapping between TDEXX and GWA-BSE is not exact
because the $\Hx^4$ terms do not have counterparts in the GWA-BSE.
These differences may indicate the inherently different nature of
TDDFT and the GWA-BSE approach.

In summary, we derived the expression of the EXX kernel for
insulators and showed that it has a long-wavelength behavior as
the exact $\Fxc$ unlike the LDA and GGA kernels. The common
conception that DFT is not suitable for the study of electronic
excitations of solids was mainly derived by adopting {\em
qualitatively} incorrect LDA and GGA potentials and kernels and
the difficulty of going beyond them. Coupled with the already
available EXX potential, we expect the numerical realization of
the EXX kernel will open up a new window of opportunity for the
first-principles study of electronic excitations in
solids.~\cite{TDEXX}

\acknowledgements We thank E. K. U. Gross for motivating us to
write up this paper, and W.  Domcke for providing research
facilities.  Y.-H. Kim also acknowledges X. Gonze for the
discussion regarding Ref. \onlinecite{Ghosez97} at Electronic
Structure Workshop 1997. This work was supported by the Humboldt
Foundation (YHK) and the Deutsche Forschungsgemeinschaft and the
Fonds der Chemischen Industrie (AG).


\newpage


\begin{thebibliography}{}
\bibitem{AU}
We use Hartree atomic units, $\hbar = e = m_e = 4\pi \epsilon_0 =
1$, throughout.
\bibitem{quasiptl}
Quasiparticles in the conventional sense that involve ($N \pm
1$)-electron excitations in contrast to ``KS quasiparticles" that
describe $N$-electron (electron-number conserving) one-particle
excitations.
\end{thebibliography}
\end{document}